\documentclass[floatfix,preprintnumbers,superscriptaddress,showpacs,
               showkeys,prb,preprint]{revtex4} 

\usepackage[version=3]{mhchem} 
\usepackage{graphicx}
\usepackage{dcolumn}
\usepackage{amsmath, amsfonts, amssymb, bm}
\usepackage{subfigure}
    
\begin{document}
\bibliographystyle{apsrev} 
\preprint{A/Fen2K6/articulo.tex}

\title{
Ab-initio study of the electronic and geometrical structure 
of tripotassium intercalated phenanthrene
} 

\author{P.L. de Andres
\footnote{pedro.deandres@csic.es}
}

\affiliation{
Instituto de Ciencia de Materiales de Madrid, CSIC, 
Cantoblanco, 28049 Madrid, Spain.
}

\author{A. Guijarro}

\affiliation{
Departamento de Quimica Organica and Instituto Universitario de Sintesis Organica,
Universidad de Alicante, San Vicente del Raspeig, 03690 Alicante, Spain.
}

\author{J.A. Verg{\'e}s}

\affiliation{
Instituto de Ciencia de Materiales de Madrid, CSIC, 
Cantoblanco, 28049 Madrid, Spain.
}

\date{\today}

\begin{abstract}
The geometrical and electronic structure of tripotassium doped
phenanthrene, \ce{K3C14H10}, have been studied by
first-principles density functional theory.
The main effect of potassium doping is to inject charge in
the narrow phenanthrene conduction band, rendering the system
metallic.
The Fermi surface for the experimental X-rays unit cell
is composed of two sheets with marked one
and two dimensional character respectively. 
\end{abstract}

\pacs{74.70.Kn,74.20.Pq,61.66.Hq,61.48.-c}


\keywords{phenanthrene, potassium, K3, C14H10, polyciclic hydrocarbon intercalation,
geometrical structure, electronic structure,fermi surface,organic crystal superconductivity}


\maketitle

\section{Introduction.}

There is much interest in the properties of
polycyclic aromatic hydrocarbons (PAHs) due to their
remarkable potential in a number of fields including
electronic devices, energy storage, molecular recognition, etc. 
Phenanthrene (\ce{C14H10}) is one of the smallest molecules
of that family with very interesting electronic properties derived
partly from its "arm-chair" edge termination, as opposed to
the "zigzag" termination characteristic of the Acene family
(e.g. anthracene, with the same molecular formula).
These two different terminations have profound effects in
in the limit of very large PAHs (i.e., graphene nanoribbons) since
anthracene like strips of material are metallic while
phenanthrene like strips can be either metallic or semiconducting
depending on their widths.\cite{nakada96,yoshizawa98}
Furthermore, organic crystals based in different PAHs 
show an amazing large number of useful properties that can be
easily tuned by the addition of appropriate contaminants. 
A recent example is the discovery of a whole new family of 
organic high-Tc superconductors when doped with alkali 
metals, as reported from experiments in picene\cite{mitsuhashi10},
phenanthrene,\cite{wang11} and most recently
coronene.\cite{submitted.coronene11}
Interestingly enough, it has been suggested that the superconducting temperature
increases with the size of the molecule forming the solid, which, if
finally proved, should make these materials of considerable interest.\cite{wang11}
In an effort to understand these challenging experimental reports,
a number of theoretical papers have investigated the nature of
pristine and doped related molecular crystals, mostly for
picene,\cite{kosugi09,hansson06,okazaki10,roth11,K6Pi2}
but also for pentacene,\cite{craciun09}
and coronene.\cite{kosugi11}
In this paper we undertake the theoretical study of 
tripotassium intercalated phenanthrene (\ce{K3C14H10})
that has been very recently reported to become
superconductor with $T_{C}=5$ K.\cite{wang11}
The mechanism for superconductivity is not clear yet,
but a dependence of $T_{C}$ with external pressure has
been found that hints to a non-conventional type of superconductor.
To elucidate these questions it is of paramount importance to gather information
about the geometrical and electronic structure of the material;
which is the main motivation for this work.

\begin{figure}
\includegraphics[clip,width=0.99\columnwidth]{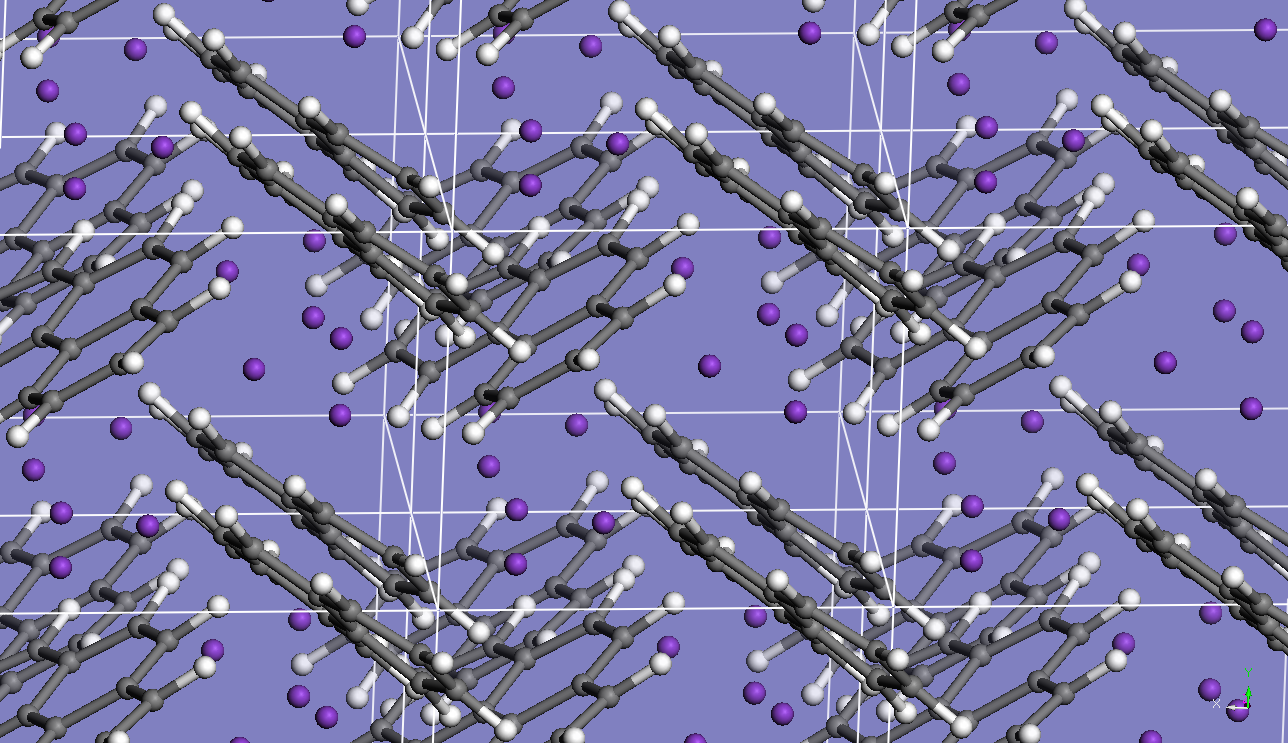} \\
\includegraphics[clip,width=0.99\columnwidth]{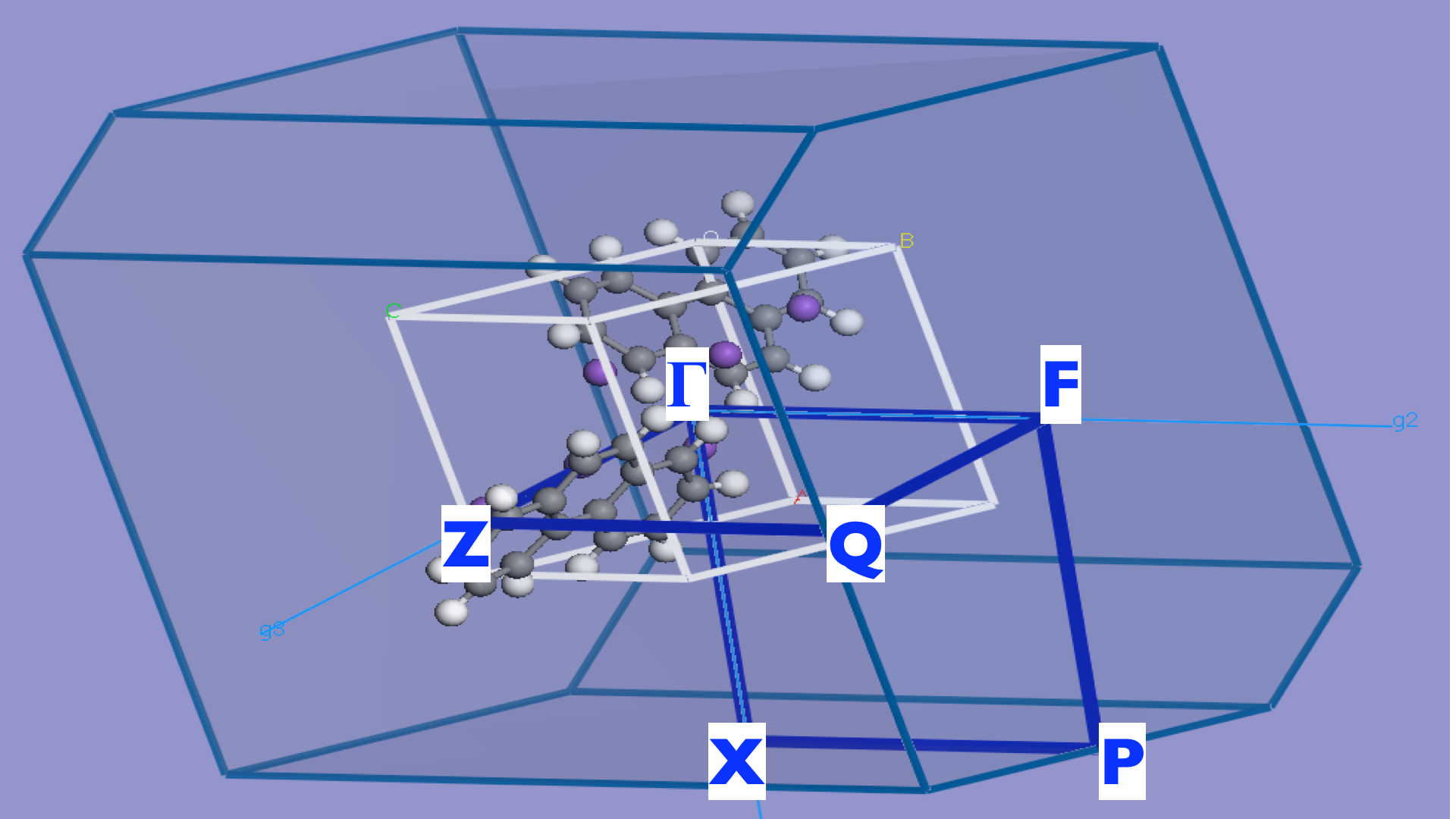}
\caption{
(a) Optimized geometry structure for \ce{K3} doped phenanthrene 
(upper pannel).
(b) Brillouin zone with the selected path for the band structure
calculation shown in Fig. \ref{fgr:BSK3} (lower pannel).
}
\label{fgr:UC}
\end{figure}

\section{Methods.}

Theoretical calculations for model systems of interest have
been performed using ab-initio
density functional theory (DFT).\cite{Hohenberg64}
Wavefunctions have been expanded in a plane-wave basis
up to a cutoff of $680$\,eV and were sampled on
a Monkhorst-Pack $5 \times 7 \times 5$ mesh inside the Brillouin zone.
Electronic bands were obtained using a smearing width of
$\eta=0.01$\,eV.
Carbon and hydrogen atoms have been described by 
accurate norm-conserving pseudopotentials.\cite{Lee96}
For the exchange and correlation (XC) potential the 
local density approximation (LDA)\cite{LDACA,LDAPZ} has been chosen
because of the following reasons:
(i) It has been successfully used to describe a large variety of carbonaceous
materials; in particular, it describes reasonably well the equilibrium configuration
of graphite (unlike GGA that fails to give a reasonable interlayer
distance), diamond, the energetic equilibrium between graphite and
diamond, and it has been successfully used in many studies of graphene layers,
nanotubes, etc. 
(ii) It yields C-C and C-H distances for many organic molecules
within -2\% error or less
(e.g., benzene, pentacene, \ce{C60}, picene,
coronene, anthracene, etc, but in particular and most relevant for this work,
phenanthrene).
(iii) It yields the lattice constant for bcc K within -5\% error, and
it brings an appropriate description for the electronic structure of 
alkali metals.
(iv) Finally, it describes reasonably well 
organic crystals of interest, in particular 
it yields the phenanthrene unit cell parameters
within -5\%. Indeed, LDA has been the approximation of 
choice in many of the ab-initio DFT studies for these systems
so far.
\cite{kosugi09,roth11,kosugi11,K6Pi2,pda08b,verges10} 
We notice that in all those examples LDA yields shorter bond lengths 
and tends to overestimate binding energies if
compared with experimental results, 
while other approaches, e.g. GGA, tend to result in larger distances
and underestimate the interaction.
Generally speaking, a local density approximation is conceptually
clear and robust, bringing a good compromise between 
accuracy and computational effort, yielding physically 
motivated conclusions.
Total energies were computed with the CASTEP
program,\cite{Clark05} as implemented in
Materials Studio.\cite{accelrys}

\section{Results.}
A monoclinic unit cell (UC) displaying a
P2$_{1}$ symmetry and including two
phenanthrene molecules in the basis
has been set up with
parameters derived from an X-rays analysis
(ref\cite{Trotter63}, Fig. \ref{fgr:UC}).
Using the formalism and the parameters described above this system can be optimized to minimize residual forces and stresses. The calculation has been considered converged when the maximum stress on the unit cell has been $\le 0.2$ GPa, the maximum force on any of the atoms on basis in the unit cell has been $\le 0.02$ eV/{\AA}, and the maximum change on the total energy has been $\le 0.0001$ eV.  The efficient BFGS method has been used to explore the huge configurational parameter space.\cite{BFGS} Convergence against the energy cutoff and the k-points mesh confirm that the enthalpy is accurate to at least $0.01$ eV/atom (fractional accuracy is $\approx 0.0001$\%); all these convergence criteria are enough for our purpose of identifying candidate local minima for the structure, and for interpreting the associated electronic structure.
The equilibrium unit cell parameters are:
$a=8.46$,
$b=6.16$,
$c=9.47$ in {\AA}, and
$\beta=97.7^{\circ}$,
$\alpha=\gamma=90^{\circ}$.
These values show with respect to the X-rays experimental determination\cite{Trotter63}
fractional errors of -5\%, -3\%, and -3\% for $a$, $b$ and $c$, respectively.
Similarly to the well-known case of graphite, a gradient generalized approximation to
the exchange and correlation brings larger errors, in particular
13\%, 8\% and 7\%, respectively.
The optimization of the unit cell to retire residual stress yields a $-0.22$ eV/unit cell
improvement in the total energy.
Most of the atomic coordinates in the phenanthrene molecules forming
the unit cell basis agree quite well with the experimental ones
(i.e. $\pm 0.05$ {\AA}, we notice
that different published experimental determinations of the crystal could
differ by $\approx \pm 0.02$ {\AA}).
The exception comes from the four H atoms participating
in the CH-$\pi$ interaction between
first-neighbors molecules; the theory to experiment difference
goes up to $0.18$ {\AA}. 
This is not at all surprising since 
we are bound to compare a theoretical calculation
performed at T=0 K with an experimental structural determination at
room temperature where the thermal vibrations should affect more to light atoms
like hydrogen, especially if they form part of a weak bond keeping 
together the two molecules in the unit cell.
Finally, the average angle between the planes containing
the two molecules stays within $2^{\circ}$
of the experimental value.
Therefore, these results for pristine phenanthrene confirm the adequacy and set the
accuracy of the proposed formalism.


Experimental analysis of \ce{K3C14H10} doped crystal
shows a contraction of the unit cell in the $\vec b$ and $\vec c$ 
directions, an expansion in the $\vec a$ direction,
and a small increase in the angle $\beta$. The unit
cell volume is overall decreased by -3.5\%.
From a theoretical point of view, however, intercalation
of three potassium atoms creates an internal stress that results in
an increase of the unit cell volume by +9\% with a
corresponding gain in energy of $-0.91$ eV/cell.
Since the interesting effects are expected to be associated
with the basic 2D character of the material
it might be interesting to notice that such a increase is not
evenly distributed in the three crystallographic directions, the 
theoretically predicted area $a \times b$ only increases by
+2\%, while the increase in $c$ is responsible for the rest.
We remark that the theoretically description in the
$c$ direction should include long-range interactions that, however,
would rend ab-initio self-consistent calculations too costly 
(this scenario would turn even worse for large molecules).
GGA cannot explain the decrease of the unit cell volume,
in fact the GGA volume increases by +37\%.
A crude hard-sphere model for K would predict an increase
of the unit cell volume by +40\% to include six interstitial K atoms
in the unit cell; this is clearly not a realistic value but
nonetheless it explains a positive increase of the volume, not a decrease.
To the best of our knowledge there is not a sensible theoretical
explanation for the reported decrease of volume after
potassium intercalation.
The best X-rays analysis for the doped material was unable to yield
a reliable structural determination with the positions of
the atoms in the unit cell basis, therefore the particular
positions of interstitial potassium, or even the exact stoichiometry in the unit cell,
has not been experimentally
resolved and the best we can do is to try different candidates
to investigate the one corresponding to the minimum total energy
and lower stresses and forces. 
According to our previous experience with the related system tripotassium-intercalated picene we have started the optimization of geometrical structures by locating potassium atoms over the hollow positions on the carbon rings forming the molecule. 
Subsequent geometrical optimization allows all the atoms to relax to the nearest local minima by using an efficient quasi-Newton method.
Taking for granted that the X-rays experimental analysis faithfully 
reproduces a decrease of the volume upon intercalation of 
potassium we have explored whether a smaller amount of interstitial
atoms could explain this fact without success.  
Therefore, our structure should represent, in the worst case,
a local metastable minimum intervening with a some weight in the appropriate 
thermodynamical average at a given T. 
Notice that in all cases the forces on all the atoms in the unit cell have been minimized as explained above.
This is the same course of action as taken before for
potassium doped picene and coronene, and it seems
a reasonable one to gather further knowledge on these 
materials.\cite{okazaki10,kosugi11}


\begin{figure}
\includegraphics[clip,width=0.99\columnwidth]{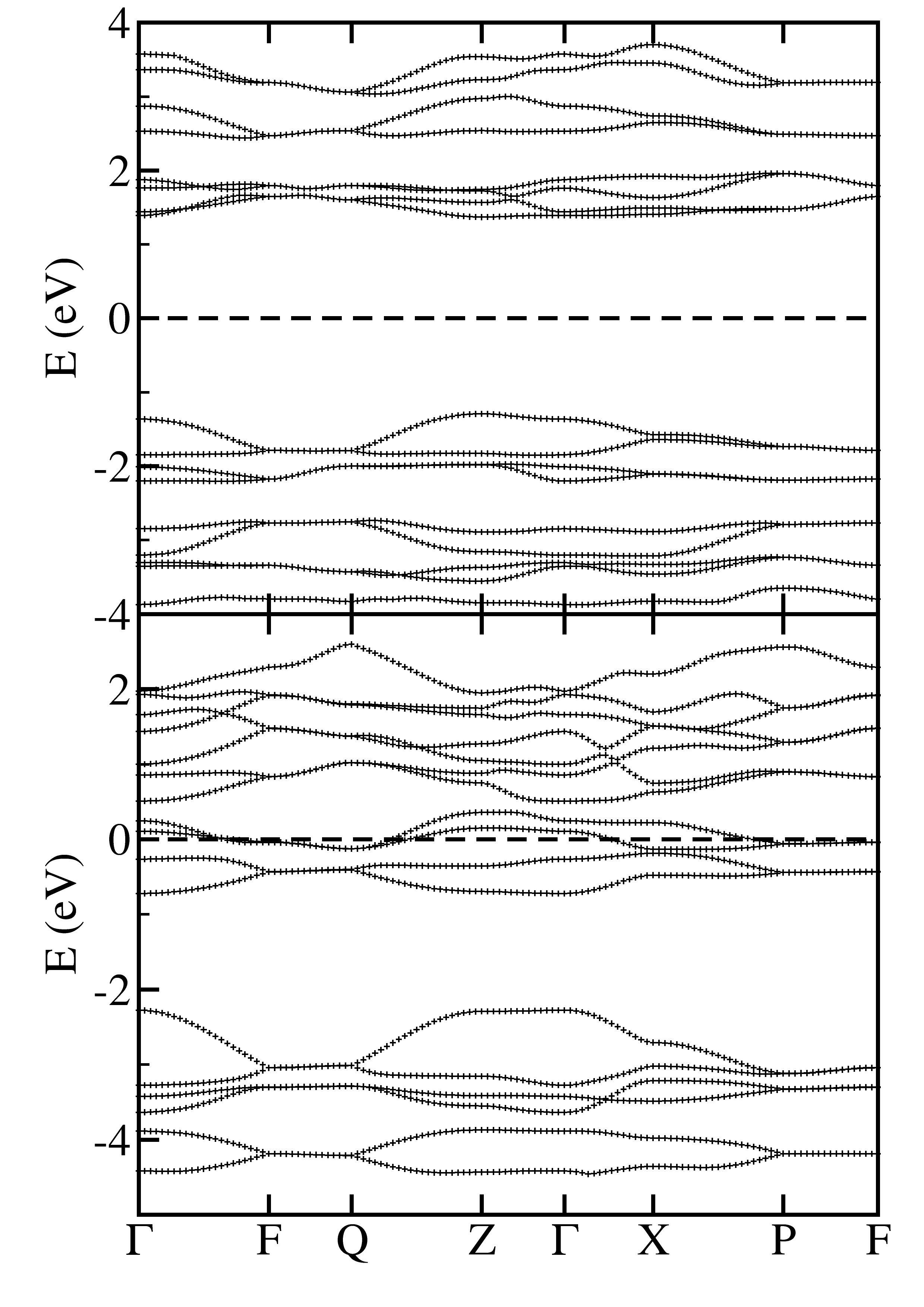} 
\caption{
Upper pannel:
Band Structure for the phenanthrene crystal.
Lower pannel: electronic structure
for the \ce{K3} doped crystal corresponding to the X-rays
structural determination\cite{wang11}.
The Fermi energy has been taken as origin for energies (dashed lines).
}
\label{fgr:BSK3}
\end{figure}

\begin{figure}
\includegraphics[clip,width=0.99\columnwidth]{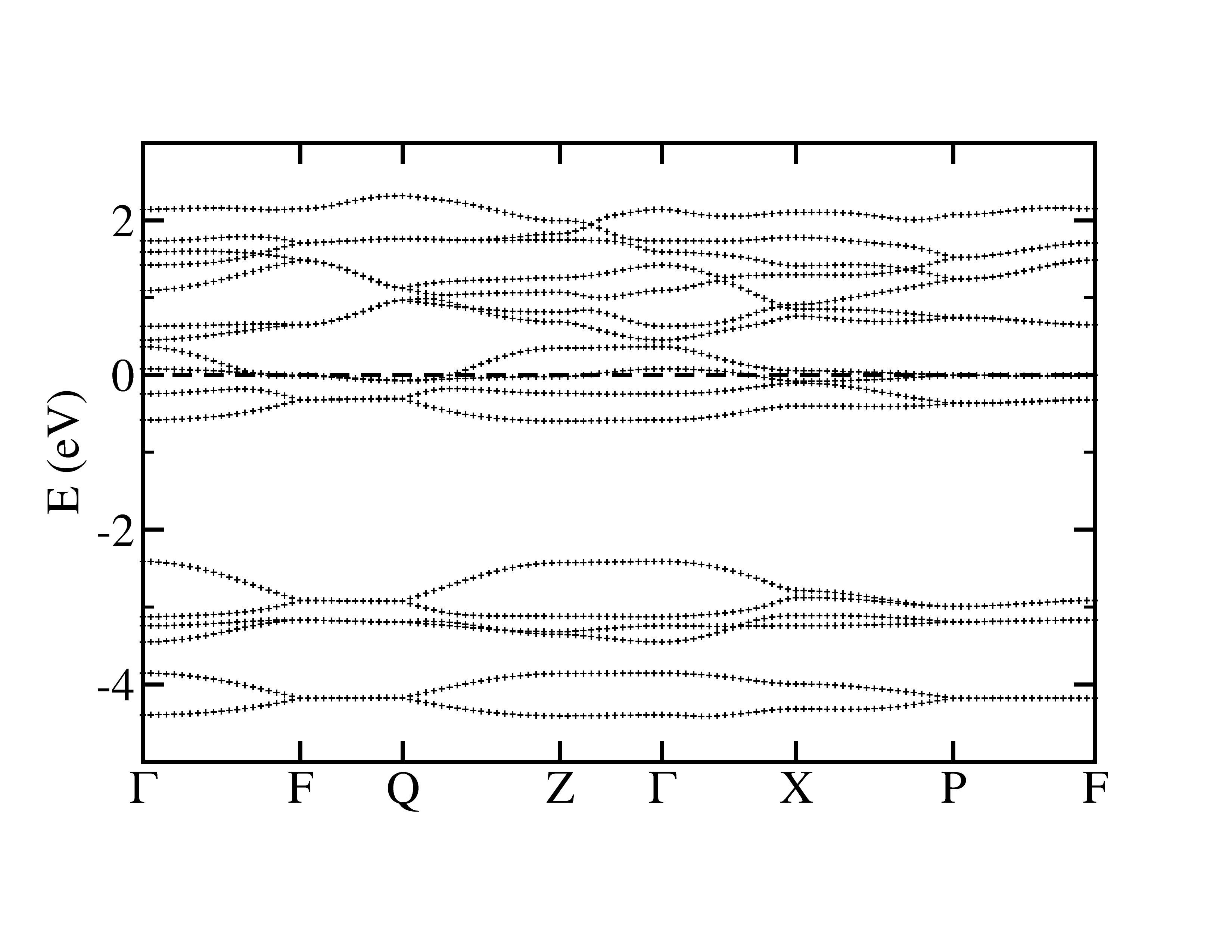}
\caption{
Band Structure for the relaxed structure of the 
\ce{K3} doped crystal where the unit cell has been optimized
to remove stress ($S_{max} \le 0.2$ GPa
(all symbols as in Fig. \ref{fgr:BSK3}).
}
\label{fgr:BSK3RLX}
\end{figure}

\begin{figure}
\includegraphics[clip,width=0.99\columnwidth]{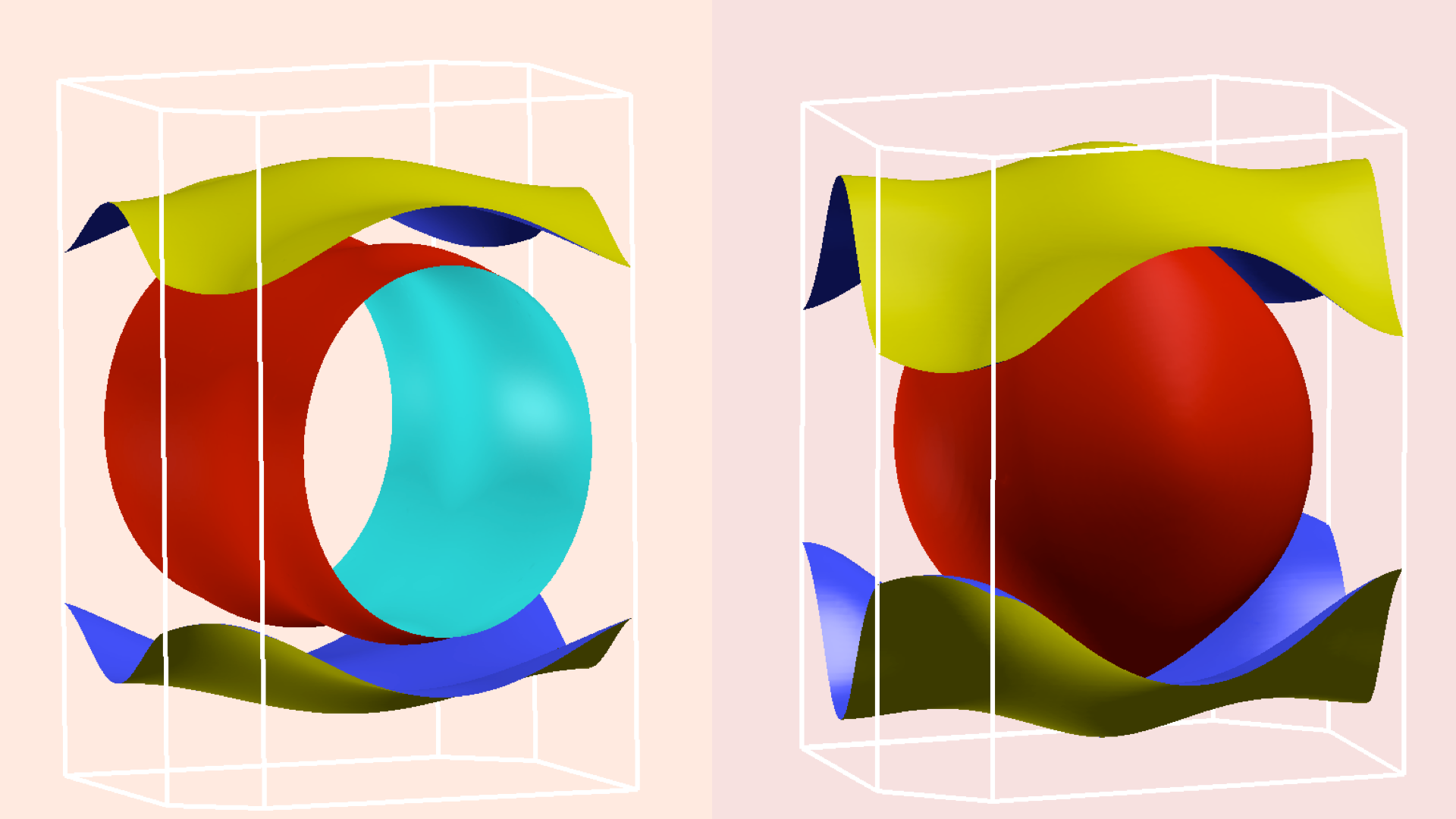} 
\caption{
For the doped phenanthrene crystal the
Fermi surface corresponding to the X-rays determination
(left pannel) is compared to the one related to the
relaxed unit cell (right pannel). 
}
\label{fgr:Fermi}
\end{figure}

We compute the electronic band structure along a selected path
on the Brillouin zone (Fig. \ref{fgr:UC}). Our 
LDA calculation 
yields for the phenanthrene crystal
a direct gap at $\Gamma$ of $2.75$ eV,
to be compared with an experimental one
of $3.16$ eV.\cite{yoshizawa98,bhatti00}
Furthermore, we compute a heat of formation for the crystal of $-1.7$ eV/cell
(with respect to a gas of single phenanthrene molecules), that may
be compared with an experimental value (STP) of $-1.83$ eV/cell.\cite{nist}
Finally, we estimate the CH-$\pi$ interaction between two molecules
as $-0.26$ eV (attractive).
For the experimental X-rays determined unit cell
the important features
in the band structure come from the four bands at the top of the valence
band and the four bands at the bottom of the conduction band. 
As it has been discussed in similar systems,
these show a strong molecular character with weak overlap,
and are related to the HOMO and HOMO-1 and LUMO and LUMO+1
molecular orbitals respectively.\cite{kosugi11}
The main effect
for the crystal intercalated with potassium, \ce{K3C14H10},
is an effective doping of the conduction band that renders the
system metallic (Fig. \ref{fgr:BSK3})
while the residual stress on the unit cell plays to slightly increase the width 
and to modify the shape of a few bands near the Fermi energy 
(Fig. \ref{fgr:BSK3RLX}); this is most clearly seen in the Fermi surface pictures below.
Our calculations show that there is an important charge transfer
from the alkali atom to the organic molecules and
the Fermi energy is located in the middle of a narrow band 
of $0.5$ eV width that
becomes responsible for the metallic properties of the 
doped material.
This band might be sensitive to correlation effects that should
be analyzed in a different framework.

The experimental unit cell used for the band structure calculation, however,
is under an approximate isotropic pressure of $3$ Gpa in the
present formalism. As the $T_{C}$ of this material
seems to be sensitive to external pressure, we have checked
the effect of removing the stress by letting the unit cell relax to
an stable equilibrium condition. As the unit cell increases slightly in volume,
it decreases the hybridization between the LUMO orbitals of both
molecules, and the conduction band crossing the Fermi energy
becomes less dispersive. However, no big changes are observed
along the chosen path to represent the bands.
The Fermi surface, on the other hand, is more sensitive to stress: 
we have compared the results for the relaxed and 
unrelaxed unit cells in Fig. \ref{fgr:Fermi}.
The ab-initio self-consistent Fermi surface has been obtained on a
denser Monkhorst-Pack
grid in the irreducible part of 
the Brillouin zone (280 k-points), and it has been drawn
using XCrysDen.\cite{xcrysden} 
A two
sheet Fermi surface has been obtained: 
For the X-rays derived UC, one of the sheets is nearly planar
and hints to the strong one-dimensional character of states
involved in the metallic conduction on this material. The other
sheet is nearly cylindrical and reveals a more two-dimensional
character (Fig. \ref{fgr:Fermi}). 
After relaxing the unit cell, the planar sheet changes very little,
but the cylindrical-like shape becomes an spheroid where the
BZ boundaries are now not touched and the corresponding gaps
occurring near the Fermi energy are closed.

\begin{figure}
\includegraphics[clip,width=0.99\columnwidth]{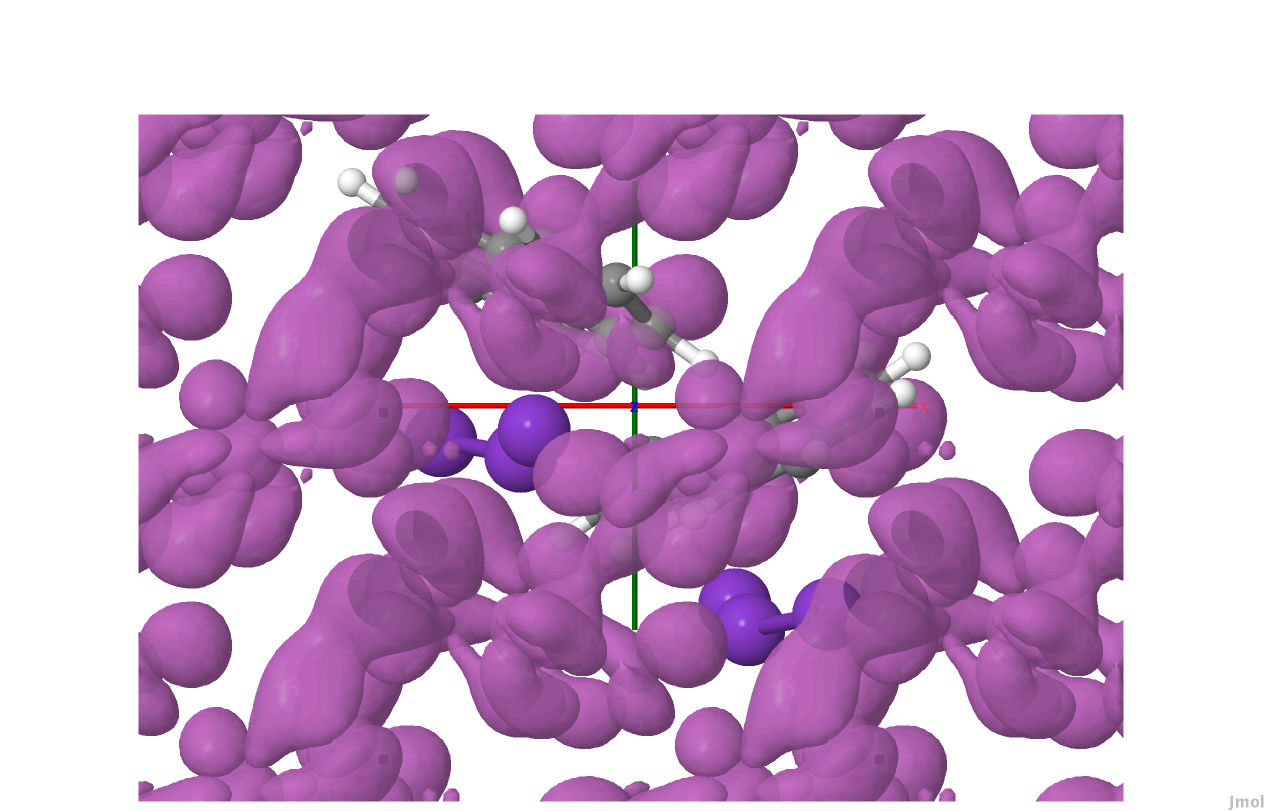}
\caption{
Iso-density corresponding to the four bands nearest the Fermi energy
inside the unrelaxed unit cell.
}
\label{fgr:isoden}
\end{figure}

The spatial distribution of electronic density\cite{rutter} confirms the above picture:
we draw in Fig. \ref{fgr:isoden} surfaces of iso-density that
integrate the four relevant bands close to the Fermi energy.
We observe the build up of extended
metallic states based in the overlap of $\pi$-like orbitals 
originated from both phenanthrene molecules. 
This is consistent with our image of an important charge transfer from
potassium to phenanthrene to build the metallic phase.


In conclusion, we have studied the geometrical and electronic structure of the clean and
potassium doped phenanthrene crystal. 
The main effect of doping is the population of the phenanthrene conduction 
band making the system metallic. 
According to our calculations the new metallic state is basically related
to properties of  phenanthrene molecular orbitals.
The experimental determination of the UC by X-rays diffraction
techniques implies stress since the volume is predicted to shrink by
$\approx 3$\% with respect to the pristine crystal, while all the
theoretical models tried have resulted in a expansion of the unit
cell volume upon potassium intercalation.
The Fermi surface under stress shows a low-dimensional character
that can be transformed by allowing the UC to deform to a global
equilibrium shape.

This work has been financed by the Spanish
MICINN (MAT2008-1497, CTQ2007-65218, CSD2007-6,
and FIS2009-08744),
DGUI of the Comunidad de Madrid (MODELICO-CM/S2009ESP-1691)
and MEC (CSD2007-41, "NANOSELECT").


\end{document}